\def\etal{{\it et al.\/}}

\def\ie{{\it i.e.\/}}

\def\Mesz{{M\`esz\`aros\/}}
 
\documentstyle[aasms4]{article}
\begin{document}

\title{Magnetospheric interactions of binary pulsars \\ as a model for
gamma-ray bursts}
\author{Mario Vietri}
\affil{ Osservatorio Astronomico di Roma \\ 00040 Monte Porzio Catone (Roma),
Italy \\ E--mail: vietri@coma.mporzio.astro.it}

\begin{abstract}
I consider a model of GRBs where they arise right before the merging 
of binary pulsars. Binary pulsars moving through the companion's 
magnetic field experience a large, motional electric field
$\vec{E}=\vec{v}\wedge\vec{B}/c$, which 
leads to the release in the pulsar magnetosphere of a pair cascade, and the 
acceleration of a wind of pure pairs. 
The energy and energy deposition rate of the wind are those of 
$\gamma$--ray bursts, provided the pulsars have a field
$\approx 10^{15}\;G$. Baryon contamination is small, and dominated by
tidal heating, leading to $M_{baryons}\approx 10^{-6}M_\odot$, as 
required by the dirty fireball model of \Mesz, Laguna and Rees. 

\end{abstract}

\section{Introduction}

The angular (Fishman 1993) and flux (Meegan \etal, 1992) distributions
of $\gamma$--ray bursts (GRBs) provide support for the suggestion 
(Usov and Chibisov 1975, Paczinsky 1986) that these sources are at 
cosmological distances. However, our current theoretical understanding 
of GRBs in the cosmological scenario presents a gap. On the one hand,
baryon--loaded fireballs  explain $\gamma$--ray emission in the optically thin 
regime, leading to non--thermal spectra, and burst durations in agreement 
with observations, provided the amount of baryon loading is suitably small, 
$10^{-6}-10^{-5}\; M_\odot$ (\Mesz, Laguna and Rees 1993). The most 
conservative scenario available at present for the release of the burst of 
energy (to be later converted into observable photons at shocks with the 
interstellar medium, \Mesz, Laguna and Rees 1993) identifies GRBs with 
phenomena occurring when compact objects merge (Eichler, Livio, Piran and 
Schramm 1989); the greatest success of this model is the prediction 
of a GRB rate in agreement with observations (Narayan, Piran, and Shemi 1991).
However, the proposed ways in which this model should release energy suffer 
from three serious defects. First, calculations by Janka and Ruffert (1996) 
and Ruffert, Janka, Takahashi and Schaefer (1996) show
that the energy release in neutrinos, which should lead to the formation of a
pure pair fireball in the models of Eichler \etal, (1989), Mochkovitch \etal, 
(1993), Woosley (1993), is less efficient than conjectured by about $2-3$ 
orders of magnitude. Second, in the magnetic mode of energy release
conjectured by Narayan, Paczinsky and Piran (1992), it is not at all obvious
whether the expected ballooning out of the magnetic field can really 
shed the baryonic component of the differentially rotating disk, 
making the desirable link with the dirty fireball model unlikely. Third,
fully general--relativistic numerical simulations show that the two neutron 
stars can collapse to black holes even before any merging actually takes 
place (Wilson, Mathews and Marronetti 1996).

These difficulties seem to indicate that the energy release occurs 
before merging. It is the aim of this paper to show how a fraction of the 
binary orbital kinetic energy prior to merging can be released into the 
electromagnetic channel, \ie, a pure pair plasma with photons; 
the amount of baryon contamination suitable for the generation of the
dirty fireball is also released. 

\section{The model}

I consider, for sake of simplicity, two identical pulsars, of mass 
$M = 1 M_\odot$, dipolar magnetic field of strength 
$B_0 = B_{15}\times 10^{15}\; G$ at the pulsar surface, 
spin period $T = t_{10}\times 10 \; s$, separated by a distance $R = R_7
\times 10^7\; cm$. During their orbits, the 
pulsars move across the lines of force of the other's magnetic field.
This induces a motional electric field $\vec{E}=\vec{v}\wedge 
\vec{B}/c$, where $B$ is the companion's field. 
This happens because tidal forces are unable (Bilstein and Cutler 
1992, \Mesz\/ and Rees 1992) to synchronize the orbital and spin motions. If 
they were, then there would be no motional $E$--field. Because of this 
relative inefficiency of tidal forces, I shall take $\vec{v}$ as due 
purely to orbital motion on a circular orbit, 
$v = 5.2\times 10^9 R_7^{-1/2}\; cm\; s^{-1}$,
and essentially neglect the pulsars' spin motions henceforth. 

The resulting motional electric field is 
\begin{equation}
\vec{E} = \frac{\vec{v}}{c}\wedge\vec{B} = 1.7\times 10^{11} R_7^{-7/2}
B_{15}\; statvolt\; cm^{-1} \vec{n}
\end{equation}
where $\vec{n}$ is a unit vector in the direction of $\vec{v}\wedge\vec{B}$.
The steep dependence upon $R_7$ comes from the dipolar nature of the 
magnetic field of the companion. 

This electric field refers to a binary in vacuo. 
Normally, pulsars are surrounded by free charges, and consequently, one
may wonder whether a configuration for the binary such that $\vec{E}\cdot
\vec{B} =0$ could be realized around each pulsar. However,
the two pulsars have different spin periods and, because of the
inefficiency of tidal forces, these in turn differ from the orbital period.
They also have different spin axes, different magnetic axes, all 
in turn different from the orbital angular momentum axis. Thus a force--free
solution when the pulsars have already penetrated so far inside each other's 
light cylinder seems very unlikely. The electric field of Eq. 1 is however
large, and it seems likely that it may be at least partially shielded,
though production of bound electron/positron pairs may limit this shielding 
(Usov and Melrose 1996). However, it will be shown later that most of the 
energy release occurs on a timescale $\approx 6\times 10^{-4}\; s$ when the two 
pulsars just touch. This timescale is only marginally longer than the 
light--crossing time of the system ($ 4R_{NS}/c \approx 1.4 \times 10^{-4} \; 
s$). It seems unlikely to me that exact field shielding can
hold over the whole pulsar magnetosphere at such a 
time of upheaval, so that I shall take Eq. 1 to apply and neglect shielding. 

Because none of the axes involved in this problem need be aligned, 
the electric field of Eq. 1 will in general have a component along the magnetic
field lines of the pulsar which feels such $E$--field. Thus free charges
are accelerated along magnetic field lines to relativistic energies,
limited by curvature radiation (Sturrock 1971). Equating the acceleration rate 
with the energy losses per unit time $\dot{\epsilon} = 2e^2c\gamma^4/3R_c^2$, 
where $R_c$ is the radius of curvature of the line of force being followed, 
I find a limiting Lorenz factor for electrons given by
\begin{equation}
\gamma_l = 1.5\times 10^{10}  \; B_{15}^{1/4} \; R_7^{-7/8} \; R_{c10}^{1/2}
\end{equation}
where $R_{c10}$ is the field line radius of curvature, in units of $10^{10}
\; cm$. This limiting Lorenz factor is achieved after linear acceleration over 
a distance
\begin{equation}
\label{dl}
d_l = 1.5\times 10^2 \; R_7^{21/8} \; B_{15}^{-3/4} \; R_{c10}^{1/2} \; cm \;.
\end{equation}
Curvature radiation from electrons/positrons with Lorenz factor $\gamma_l$
produce photons with typical energies
\begin{equation}
h\nu_l = \frac{3 c h \gamma_l^3}{4\pi R_c} = 1.0\times 10^{16} \; B_{15}^{3/4}
\; R_{c10}^{1/2} \; R_7^{-21/8} \; eV\;.
\end{equation}
The mean free path of these photons before decay into electron/positron pairs
in the given magnetic field is (Erber 1966)
\begin{equation}
\label{db}
d_B = 6.0\times10^{-3} \; B_{15}^{-7/4} \; R_{c10}^{1/2} \; R_7^{21/8} \;cm\ ,
\end{equation}
where dependence upon a slowly varying logarithmic factor has been omitted. 
These quantities show that a pair cascade begins, 
like in normal models of pulsars (Ruderman and Sutherland 1975, 
Arons 1983), even for weaker magnetic fields, or large
screening factors (both correspond to $B_{15} \ll 1$). Comptonization in the 
soft thermal photons bath emitted by the pulsar surface reduces $\gamma_l$, but
the main limitation comes from Comptonization of photons self--consistently 
emitted in the acceleration (Usov 1992), to be discussed later. 

Where does the energy for the cascade come from? In a normal diode, the 
avalanche rush of free charges to the anode occurs at the expense of the energy density of the electric field, which is thus shorted out. Here, the work done
at a rate $\vec{j}\cdot\vec{E}$, where $\vec{j}$ is the induced current density,
must occur at the expense of the motional field, Eq. 1. It is apparent that 
the appropriate source of energy to be tapped is the pulsar orbital kinetic 
energy. To see this consider the force acting on the plasma 
thusly created, $\vec{j}\wedge\vec{B}_1/c$, where $B_1$ is the field across
which the pulsar moves. Plasma moves along the lines of force of the 
pulsar from which it originated, being constrained by highly efficient
synchrotron losses; it follows that it imparts to the emitting pulsar the 
same impulse, and that the current density is 
$\vec{j} \propto sign(\vec{E}\cdot\vec{B}_2) \vec{B}_2$, where $B_2$ is the
magnetic field of the emitting pulsar. Since
$\vec{E} = \vec{v}\wedge\vec{B}_1/c$, we have $\vec{E}\cdot\vec{B}_2 = 
\vec{v}\cdot\vec{B}_1\wedge\vec{B}_2/c$. The force component along $\vec{v}$ 
then is $\vec{v}\cdot\vec{j}\wedge\vec{B}_1 = sign(\vec{v}
\cdot\vec{B}_1\wedge\vec{B}_2) \vec{v}\cdot\vec{B}_2\wedge\vec{B}_1 < 0$.
By symmetry, the total force on the pulsar is 
$\propto \pm \vec{\mu}_2\wedge\vec{B}_1$,
where $\vec{\mu}_2$ is the emitting pulsar's magnetic moment. 
Since $\vec{\mu}_2$ and $\vec{B}_1$ are independent vectors, 
the emitting pulsar may acquire a small velocity perpendicular to $\vec{v}$.

The quantities derived above (Eq. \ref{dl}-\ref{db}) suggest that a dense, 
quasi--neutral pair plasma forms. This very important assumption will be 
justified {\it a posteriori}. The quasi neutral plasma will produce a wind
with luminosity $L_{EM}$ determined by the following condition.
Plasma acceleration saturates when the plasma energy density becomes
comparable to the energy density of the magnetic field of the companion:
at this point the plasma becomes capable of convecting this
field, their relative motion goes to zero, the electric 
field of Eq. 1 becomes null, and no further acceleration can take place. 
This condition is 
\begin{equation}
\frac{B^2}{8\pi} = \frac{B_0^2}{8\pi} \left(\frac{R_{NS}}{R}\right)^6
= \frac{L_{EM} }{4\pi R_o^2 c }
\end{equation}
where $R_{NS}= 10^6\; cm$ is the neutron star radius, and $R_o$
is the radius of the magnetosphere within which the
energy release occurs. The above condition can be rewritten as
\begin{equation}
\label{lem}
L_{EM} = \frac{1}{2} B_0^2 \left(\frac{R_{NS}}{R}\right)^6 R_o^2 c = 
10^{52} B_{15}^2 \left(\frac{R_7}{0.2}\right)^{-6}\; erg \; s^{-1}\; ,
\end{equation}
scaled to the moment when the pulsars are just about to merge, with a small 
correction accounting for the $\vec{B}_1\cdot\vec{B}_2$ term, and with
$R_o = 3 R_{NS}$ because, for larger values,
the magnetic field energy decreases very fast. 
The above luminosity is most likely self--regulating. 

The total luminosity released by gravitational wave radiation (Shapiro and 
Teukolsky 1983), $L_{GW}\propto R_7^{-5}$, much exceeds the electromagnetic 
luminosity:
\begin{equation}
\label{ratio}
\frac{L_{EM}}{L_{GW}} = 0.012 B_{15}^2 \left(\frac{1 
M_\odot}{M_{NS}}\right)^5 \left(\frac{0.2}{R_7}\right)\; .
\end{equation}
Thus the orbital decay is still dominated by gravitational radiation, for 
which $R_7 \approx 0.2 (t/t_0)^{1/4}$, with $t_0 = 6\times 10^{-4}\; s$. Then
from Eq. \ref{lem} I find $L_{EM} \propto (t/t_0)^{-3/2}$: 
the wind actually increases fast enough to look like a burst.
Integration of Eq. \ref{lem} over an orbital decay dominated by gravitational
radiation yields for the total energy release
\begin{equation}
\label{eem}
\bigtriangleup E_{EM} = 8\times 10^{50} B_{15}^2
\left(\frac{2R_{NS}}{R_f}\right)^2
\; erg
\end{equation}
where $R_f$, the radius at which the integration has been stopped, has been 
taken at the time when the pulsars barely touch, leading to good agreement with
observations for which $E = 4\times 10^{51}\; erg$ (Piran 1992) for $B_{15}
\approx 2$. 

I now examine the assumption of a quasi--neutral plasma. The hyperrelativistic
particles described above (Eq. \ref{dl}-\ref{db}) lose energy as $\gamma$-ray
radiation, which is transformed into $e^-/e^+$ pairs and photons, with the help 
of the decay process $\gamma+B \rightarrow \gamma'+\gamma''+B$ (Adler 1971). In 
this environment, the pairs' Lorenz factor is limited by interaction with 
photons: Usov (1992) finds $\gamma_p \approx 10$ and the total optical depth of 
the wind for $e^+/e^-$ pair at the moment of pulsars' contact is (Usov 1992):
\begin{equation}
\tau \approx 10^{14} B_{15}^2 \gamma_p^{-2}
\end{equation}
It can also be shown that the wind is optically thick to Compton scattering, 
$\gamma$--ray absorption and photon decay. This
implies that charges of both signs are convected away by photon
viscosity; this contrasts with normal pulsars,
where positrons and electrons, which are not coupled by radiative
processes, freely stream in opposite directions. The remaining currents
can be compared with the Goldreich and Julian maximum currents 
(Ruderman and Sutherland 1975, Fawley, Arons and Scharlemann 1977) 
$I_{GJ} \approx 4\pi R_{NS}^2 c E /r$,
where $r$ is a typical distance over which the current flows, which will
drop out of the result.
An estimate is provided by the fact that the term $\vec{j}\wedge\vec{B}_1/c$
in the plasma momentum equation must yield the pulsar deceleration, 
$L_{EM}/v$. I find $\int j \; d\!V \approx L_{EM} c/ v B$; since
$\int j\; d\!V \approx I r/3$, I find, using Eqs. \ref{lem}, and 1 
\begin{equation}
\frac{I}{I_{GJ}} = \frac{3}{4\pi}
\left(\frac{c}{v}\right)^2
\left(\frac{R_o}{R_{NS}}\right)^2
\left(\frac{R_{NS}}{R}\right)^6 \approx \frac{1}{4}
\left(\frac{2 R_{NS}}{R}\right)^5 \;.
\end{equation}
This shows that the currents required to decelerate the pulsar 
in this model do not vastly exceed the maximum currents that can be
supported by the system without disruption of the electromagnetic
configuration assumed. This justifies the assumption of a quasi--neutral
plasma, and supports the correctness of the total energy release
computed in Eq. \ref{eem}.

Since the burst takes place in the magnetosphere, essentially no baryons
will be involved. There are however three major sources of contamination.
A fraction $\tau^{-1}$ of the total luminosity will hit the pulsar surface,
causing a heating. Even if all of this luminosity were used to unbind 
baryons from the surface, rather than be converted into heat, it would still
amount just to $\la 10^{-9} M_\odot$, negligible with respect to the
demands of the dirty fireball model. Thermionic emission is negligible for
surface temperatures $ \la T_i = 3.5\times 10^5 (B/10^{12} G)^{0.73}\; K$, 
which becomes very large for $B_{15} \approx 2$ (Usov and Melrose 1995).
However \Mesz\/ and Rees (1992) showed that tidal heating is capable of 
releasing a baryon mass which grows, as the separation decreases, as
\begin{equation}
\bigtriangleup\!M = 10^{-6} \left(\frac{0.2}{R_7}\right)^{10} M_\odot\; ;
\end{equation}
for $R_7 = 0.2$, this is exactly the ideal value demanded by the dirty fireball 
model (\Mesz, Laguna and Rees 1993). These computations seem to assume 
a viscosity that is too high by several orders of magnitude (Reisenegger
and Goldreich 1994, Lai 1994). Yet, even these more sophisticated
computations imply that 
$\approx 10^{-6}$ of the binary orbital energy can go into tidal heating; if
then a tidal wind outflow is one of the leading energy loss mechanisms, 
possibly in near equipartition with the others, still $\approx 10^{-6}
\; M_\odot$ can be carried away by the tidal wind. 
Thus magnetospheric release of the burst energy turns baryon contamination
by tidal heating from the liability it was (it makes contact with the
dirty fireball more difficult) for the models of 
Eichler \etal\/ (1989) and Narayan, Paczinsky and Piran (1992), into an
asset.

\section{Discussion}

Since the photons to be observed are released in a later shock, this model
makes no testable predictions upon the spectral properties of GRBs, for 
which one is referred to the dirty fireball (\Mesz, Rees and Papathanassiou 
1994). However, the present model predicts that, since the force on the 
emitting pulsar is not necessarily parallel to the orbital velocity, an orbital 
eccentricity can be induced on the orbit, previously circularized by 
gravitational radiation. The amplitude of this eccentricity is 
\begin{equation}
e = \frac{r_a-r_p}{r_a+r_p} \approx \delta\!e \approx \sqrt{2} 
\frac{\delta\!h^2}{h^2} 
\approx \left(\frac{2 L_{EM} T_o} {|{\cal E}|}\right)^{1/2} \approx 0.07
\end{equation}
where $r_a, r_p, {\cal E}, h, T_o$ are the orbit's apoastron, periastron,
energy, angular momentum, and period, and the second and third equalities hold
for nearly circular orbits. Since circularization occurs on the gravitational
radiation timescale which is longer than the orbital period, an eccentric orbit 
will persist for a few passages. The ensuing electromagnetic luminosity
$\propto R^{-6}$ (Eq. \ref{lem}), and thus eccentricities of $\approx 0.07$
suffice to yield variations of luminosity between periastron and apoastron of
order $\delta\!L_{EM}/L_{EM} \approx 1$, on the orbital timescale at the moment 
of merging, $\approx 10^{-3}\; s$. Relativistic effects imply that, at the
moment of forming the shock that will radiate the observable photons, 
this timescale has become $\approx 1\; s$; both the amplitude variations
and the timescale satisfy the requirements of the inital, `seed' oscillations
postulated by Stern and Svensson (1996) to give rise, through the pulse
avalanche model, to the whole range of bursts' temporal structure (Fishman
and Meegan 1995). 

This model requires (Eq. \ref{eem}) that the two pulsars have a large magnetic 
field, possibly either from birth (Thompson and Duncan 1993) 
or upon undergoing the transition from a white dwarf
to a neutron star (Usov 1992) after accretion from a normal companion.
Alternatively and more attractively, the $B$--field may be dynamically 
generated: \Mesz\/ and Rees (1992) suggested that the magnetic field is wound 
and sheared by the wind that carries away the baryons blown away by tidal 
heating. The ensuing equipartition field they derive (see discussion after 
their Eq. 4.1) is
\begin{equation}
B_{eq} = 5\times 10^{14}\left(\frac{0.2}{R_7}\right)^{7} \; G
\end{equation}
close to that required by Eq. \ref{eem}, even though very uncertain because of 
the very steep dependence upon $R_7$. It thus seems possible that dynamo action
powered by tidal heating generates large magnetic fields, and these in turn,
through the process described in this paper, lead to GRBs. A full study of this
effect is in preparation. 

In short, I have discussed a way to transform the binary pulsars' orbital
kinetic energy into an electromagnetic channel; this model is not based upon 
any exotic physics, particle acceleration is unavoidable, and it links smoothly 
with the work of \Mesz, Laguna and Rees (1993) and Narayan, Piran and Shemi 
(1991), thus providing a coherent view of GRBs. 

\acknowledgments

It is a pleasure to acknowledge interesting conversations with A. Cavaliere, 
M. Salvati and L. Stella, and helpful comments from an anonymous referee.

\end{document}